\documentclass[reqno]{amsart}

\usepackage{booktabs}
\usepackage{IEEEtrantools}
\usepackage{amssymb,latexsym,amsfonts,amsmath}
\usepackage{mathrsfs}
\usepackage{graphicx}
\usepackage{dsfont}
\usepackage{tikz}
\usetikzlibrary{calc,shapes,arrows}

\usepackage{algorithm}
\usepackage{algorithmic}

\usepackage{xspace}

\newtheorem{theorem}{Theorem}[section]
\newtheorem{lemma}[theorem]{Lemma}

\newtheorem{definition}[theorem]{Definition}

\newtheorem{remark}[theorem]{Remark}
\newtheorem{assumption}{Assumption}
\numberwithin{equation}{section}

\newcommand{\R}{{\mathbb{R}}}

\newcommand{\N}{{\mathbb{N}}}

\newcommand{\Let}{:=}
\newcommand{\EE}{\mathds{E}}
\newcommand{\PP}{\mathds{P}}

\usepackage{fancyhdr}

\newenvironment{nouppercase}{%
	\renewcommand{\uppercasenonmath}[1]{}}{}

\begin{document}

\begin{abstract}
In this work, we propose a compositional scheme for the safety controller synthesis of interconnected discrete-time stochastic systems with Markovian switching signals. Our proposed approach is based on a notion of so-called \emph{control storage certificates} computed for individual subsystems, by leveraging which, one can synthesize state-feedback controllers for interconnected systems to enforce safety specifications over finite time horizons. To do so, we employ a sum-of-squares (SOS) optimization
approach to search for \emph{multiple} storage certificates of each switching
subsystem while synthesizing its corresponding safety controller. We then utilize dissipativity theory
to compositionally construct barrier certificates
for interconnected systems based on
storage certificates of individual subsystems. The proposed dissipativity-type compositional conditions can leverage the structure of the interconnection topology and be fulfilled independently of the number or gains of subsystems. We eventually employ the constructed barrier
certificate and quantify upper bounds on the probability
that the interconnected system reaches certain unsafe
regions in a finite time horizon. We apply our results to a room temperature network of $200$ rooms with Markovian switching signals while accepting multiple storage certificates. We compositionally synthesize safety controllers to maintain the temperature of each room in a comfort zone for a bounded time horizon.
\end{abstract}

\title{{\LARGE Compositional Controller Synthesis for Interconnected Stochastic Systems with Markovian Switching}$^*$\footnote[1]{$^*$This work was supported in part by the Swiss National Science Foundation under NCCR Automation, grant agreement 51NF40-180545.}}

\author{{\bf {\large Abolfazl Lavaei}}}
\author{{\bf {\large Emilio Frazzoli}}\\
	{\normalfont Institute for Dynamic Systems and Control, ETH Zurich, Switzerland}\\
\texttt{\{alavaei,efrazzoli\}@ethz.ch}}

\pagestyle{fancy}
\lhead{}
\rhead{}
  \fancyhead[OL]{Abolfazl Lavaei and Emilio Frazzoli}% Author on Odd page, Centred

  \fancyhead[EL]{Compositional Controller Synthesis for Interconnected Stochastic Systems with Markovian Switching} % Title on Even page, Centred
  \rhead{\thepage}
 \cfoot{}
 
\begin{nouppercase}
	\maketitle
\end{nouppercase}

\section{Introduction}

Large-scale stochastic switching systems have been considering as an important modeling framework to describe a broad range of safety-critical applications including (air) traffic networks, autonomous vehicles, delivery drones, vehicle platooning, robotic networks, etc., to name a few. Synthesizing automated controllers for this type of complex systems to enforce some high-level logic properties, \emph{e.g.,} linear temporal logic (LTL) formulae~\cite{pnueli1977temporal}, is inherently challenging, mainly due to (i) the stochasticity inside dynamics, (ii) high dimensionality, and (iii) the random behavior of switching signals. 

To deal with the encountered complexity, (in)finite abstractions have been introduced as a promising tool in the controller synthesis procedure (\emph{e.g.,}~\cite{julius2009approximations,APLS08,zamani2015symbolic,zamani2014symbolic,tmka2013}). However, the proposed abstractions-based techniques  rely on the discretization of state and input sets, and consequently, they suffer severely from the \emph{curse of dimensionality} especially when dealing with large-scale dynamical systems. Then compositional approaches for the construction of (in)finite abstractions for complex systems based on abstractions of smaller subsystems have been proposed in the relevant literature (\emph{e.g.,}~\cite{SAM17,hahn2013compositional,lavaei2018CDCJ,lavaei2019HSCC_J,lavaei2018ADHSJ,Lavaei_TAC2022,Lavaei_Survey,AmyJournal2020}).

The proposed compositional scheme in the setting of (in)finite abstractions can mitigate the effects of the state-explosion problem; however, the curse of dimensionality may still exist in the level of subsystems given
the range of state and input sets and their quantization parameters. Then  \emph{control barrier certificates} as a \emph{discretization-free} approach have been proposed for formal verification and analysis of complex dynamical systems. More precisely, control barrier certificates are Lyapunov-like functions defined over the state space of the system to enforce some conditions on both the function itself and the one-step transition of the system. Starting from a given set of initial conditions, an appropriate level set of a barrier certificate can separate an unsafe region from all system trajectories. 
Consequently, the existence of such a function provides a formal probabilistic certificate for the safety of the system. 

There have been some results, proposed in the past decade, on the verification and controller synthesis of stochastic systems via control barrier certificates. Discretization-free techniques based on barrier certificates for (stochastic) hybrid systems are initially proposed in~\cite{prajna2004safety,wieland2007constructive,prajna2007framework}. Stochastic safety verification using barrier certificates for switched diffusion processes and stochastic hybrid systems is, respectively, proposed in~\cite{wisniewski2017stochastic} and~\cite{huang2017probabilistic}. Temporal logic controller synthesis of stochastic systems via control barrier certificates is proposed in~\cite{Pushpak2019}. A controller synthesis framework  for stochastic systems based on control barrier functions is presented in~\cite{clark2019control}. Verification and control for finite-time safety of stochastic systems via barrier functions are discussed in~\cite{santoyo2019verification}. Verification of uncertain partially-observable Markov decision processes (POMDPs) with uncertain transition and/or observation probabilities using barrier certificates is discussed in~\cite{ahmadi2019safe}. Compositional construction of safety controllers for networks of continuous-space POMDPs using
control barrier certificates is recently proposed in~\cite{Niloofar_TNCS_2022}.

An introduction and overview of relevant work on control barrier functions and their application to verify and enforce safety properties in the context of safety-critical controllers are presented in~\cite{ames2019control}. Compositional construction of control barrier certificates for stochastic discrete- and continuous-time systems is, respectively, presented in~\cite{LavaeiIFAC2020} and~\cite{AmyIFAC12020}. In comparison with the current work, the proposed results in~\cite{LavaeiIFAC2020,AmyIFAC12020} only handle stochastic \emph{control} systems, while we enlarge here the class of systems to stochastic \emph{switching} system with Markovian switching signals. Compositional construction of control barrier certificates for discrete-time stochastic switched systems is studied in~\cite{Amy_LCSS20}. Although the proposed results in~\cite{Amy_LCSS20} are about stochastic switched systems, their switching signals are deterministic that are served as control inputs. In comparison, switching signals in our work are not control inputs and are randomly changing. As a result, the controller synthesis problem here is more challenging compared to~\cite{Amy_LCSS20} since it deals with two different sources of adversary inputs: (i) disturbances as the effects of other subsystems, and (ii) switching signals which are randomly changing in a finite set of modes.

Our main contribution in this work is to propose, for the first time, a compositional framework based on dissipativity theory for the safety controller synthesis of interconnected stochastic switching systems with \emph{Markovian switching} signals admitting \emph{multiple} storage certificates. To this end, we first introduce notions of stochastic storage and barrier certificates for, respectively, stochastic switching subsystems and interconnected systems. We employ sum-of-squares (SOS) optimization problems to search for control storage certificates of each individual subsystem while synthesizing its corresponding safety controller. We then compositionally construct barrier certificates for interconnected systems based on storage certificates of individual subsystems by leveraging some dissipativity-type compositional conditions. We show that the proposed compositionality conditions can utilize the structure of the interconnection topology and be satisfied independently of the number or gains of
subsystems (cf. Remark~\ref{compositionality remark} and the case study). Given the constructed barrier certificate, we quantify upper bounds on the probability that the interconnected system reaches certain unsafe regions in finite time horizons. Proofs of all statements are omitted due to space limitations.

\section{Discrete-Time Stochastic Switching Systems}\label{Sec:dt-SS}

\subsection{Preliminaries}

The probability space, in this work, is considered as $(\Omega, \mathcal{F}_\Omega, \PP_\Omega)$, where $\Omega$ is the sample space, $\mathcal{F}_\Omega$ is a sigma-algebra on $\Omega$ consisting subsets of $\Omega$ as events, and $\PP_\Omega$ is the probability measure that assigns probability to those events. Random variables are assumed to be measurable functions of the form $X:(\Omega,\mathcal{F}_\Omega) \rightarrow (S_X,\mathcal{F}_X)$. Any random variable $X$ induces a probability measure on $(S_X,\mathcal{F}_X)$ as $Prob\{A\} = \mathbb{P}_\Omega\{X^{-1}(A)\}$ for any $A \in \mathcal{F}_X.$ The topological space $S$ is a Borel space if it is homeomorphic to a Borel subset of a Polish space, \textit{i.e.}, a separable and completely metrizable space. The Borel sigma-algebra generated from Borel space $S$ is denoted by $\mathbb{B}(S)$ and the map $f: S \rightarrow Y$ is measurable whenever it is Borel measurable.

\subsection{Notation}

We employ $\mathbb{R},\mathbb{R}_{>0}$, and $\mathbb{R}_{\geq 0}$ to denote the set of real, positive and non-negative real numbers, respectively, while $\mathbb{R}^n$ represents a real space of the dimension $n$. The set of non-negative and positive integers are denoted by $\mathbb{N} := \{0,1,\ldots\}$ and $\mathbb{N}_{\geq 1}=\{1,2,\ldots\}$, respectively. Given $N$ vectors $x_i \in \mathbb{R}^{n_i}$, we use $x=[x_1;\ldots;x_N]$ to denote the corresponding column vector of the dimension $\sum_i n_i$.
The identity matrix in $\mathbb R^{n\times{n}}$ is denoted by $\mathds{I}_n$. Given functions $f_i:X_i\rightarrow Y_i$, for any $i\in\{1,\ldots,N\}$, their Cartesian product $\prod_{i=1}^{N}f_i:\prod_{i=1}^{N}X_i\rightarrow\prod_{i=1}^{N}Y_i$ is defined as $(\prod_{i=1}^{N}f_i)(x_1,\ldots,x_N)=[f_1(x_1);\ldots;f_N(x_N)]$. 

\subsection{Discrete-Time Stochastic Switching Systems} \label{dtscs}

In this work, we consider discrete-time stochastic switching systems as formalized in the following definition.

\begin{definition} \label{def-dtSCS}
	A \emph{discrete-time stochastic switching system} (dt-SS) is a tuple 
	\begin{equation}\label{eq:dt-SCS}
		\Sigma=(X,U,W,P, \mathcal P, \varsigma,F,Y,h),
	\end{equation}
	where:
	\begin{itemize}
		\item
		$X\subseteq \mathbb R^n$ is a Borel space as a state space of the
		system;
		\item $U\subseteq \mathbb R^{\bar m}$ is a Borel space as an input space of the system; 
		\item $W\subseteq \mathbb R^{\bar p}$ is a Borel space as a disturbance space of the system; 
		\item $P = \{1,\dots, m \}$  is the finite set of modes;
		\item $\mathcal{P}$ is a subset of $\mathcal{S}(\mathbb N,P)$ which denotes the set of functions from $\mathbb N$ to $P$;
		\item
		$\varsigma$ is a sequence of independent and identically distributed
		(i.i.d.) random variables from the sample space $\Omega$ to the measurable space $(\mathcal{V}_\varsigma,\mathcal F_\varsigma)$, \emph{i.e.,}
		\begin{equation*}
			\varsigma:=\{\varsigma(k):(\Omega,\mathcal F_\Omega)\rightarrow (\mathcal{V}_\varsigma,\mathcal F_\varsigma),\,\,k\in\N\}; 
		\end{equation*}
		\item $F = \{f_1,\dots, f_m \}$ is a collection of vector fields indexed by $p$. For all $p\in P$, the map $f_p:X\times U \times W\times \mathcal{V}_{\varsigma} \rightarrow X$ is a measurable function characterizing the state evolution of the system;
		\item 
		$Y\subseteq \mathbb{R}^{q}$ is a Borel space as an output space of the system;
		\item 
		$h: X \rightarrow Y$ is a measurable function that maps a state $x \in X$ to its disturbance output $y=h(x)$.
	\end{itemize}
\end{definition} 
We associate sets $\mathcal{U}$ and $\mathcal{W}$ to, respectively, sets $U$ and $W$ as collections of input and disturbance sequences $\{\nu(k):\Omega\rightarrow U,\,\,k\in\mathbb N\}$ and $\{w(k):\Omega\rightarrow W,\,\,k\in\mathbb N\}$. Both $\nu(k)$ and $w(k)$ are independent from the random variable $\varsigma(z)$ for all $k,z \in \mathbb{N}$ and $z \ge k$. 
The state evolution of dt-SS $\Sigma$ for a given initial state $x(0)\in
X$, an input sequence $\{\nu(k):\Omega\rightarrow U,\,\,k\in\mathbb N\}$, a disturbance sequence $\{w(k):\Omega\rightarrow W,\,\,k\in\mathbb N\}$, and a switching signal $\bold{p}(k):\mathbb N \rightarrow P$, is characterized by:
\begin{equation}\label{Eq_1a}
	\Sigma\!:\left\{\hspace{-2mm}\begin{array}{l}x(k+1)=f_{\bold{p}(k)}(x(k),\nu(k),w(k),\varsigma(k)),\\
		y(k)=h(x(k)),\\
	\end{array}\right.
\end{equation} 
for any $k\in\mathbb N$. For any $p\in P$, we use $\Sigma_p$ to refer to system~\eqref{Eq_1a} with a constant switching signal $\bold{p}(k) = p$ for all $k\in\mathbb N$.  
For a given initial state $ a \in X$, $\nu(\cdot) \in \mathcal{U}$ and $w(\cdot) \in \mathcal{W}$, and $\bold{p}(k):\mathbb N \rightarrow P$, a random sequence $x_{a\nu w}^p:\Omega \times\mathbb N \rightarrow X$ denotes the \emph{solution process} of $\Sigma$ under the influence of the input $\nu$, the disturbance $w$, and the switching signal $p$, started from the initial state $a$.

Given the dt-SS in~\eqref{Eq_1a} with $p,p'\in P$, the transition probability between modes is described using the following Markovian switching:
\begin{align}\label{Markovian}
	\PP\Big\{\bold{p}(k+1) = p' \,\big |\, \bold{p}(k) = p \Big\}\!=\pi_{pp'},
\end{align}
where $\pi_{pp'} \geq 0$ for any $p,p'\in P$, and $\sum_{p'=1}^m \pi_{pp'} = 1$. Accordingly, the
transition probability matrix is defined by
\begin{align}\notag
	\pi=\begin{bmatrix}\pi_{11} & \pi_{12} & \cdots  & \pi_{1m}  \\  \pi_{21}  & \pi_{22} & \cdots & \pi_{2m} \\ \vdots &  & \ddots &  \\ \pi_{m1} & \pi_{m2} & \cdots  & \pi_{mm}\end{bmatrix}\!\!.
\end{align}
The Markovian switching in~\eqref{Markovian} implies that the switching between different modes is governed by a discrete-time Markov chain with the transition probability matrix $\pi$~\cite{puterman2014markov}. We assume that the controller has an access to switching modes which is a standard assumption in the relevant literature~\cite{du2021certainty}. In particular, it is supposed that there is a mode detection device which is capable of identifying the system mode in real time so that the controller can switch to the matched mode.

In this work, we are ultimately interested in studying interconnected dt-SS without disturbances that results from the interconnection of dt-SS having disturbance signals. Hence, we consider dt-SS in~\eqref{Eq_1a} as individual subsystems and provide a formal definition of interconnected dt-SS as the following. 

\begin{definition}
	Suppose we are given $N \in \mathbb{N}_{\geq 1}$ switching subsystems $\Sigma_i = (X_i,U_i,W_i,P_i, \mathcal P_i,\varsigma_i,F_i,Y_{i},h_{i}), i\in \{1,\dots,N\}$, where $X_i \in \mathbb{R}^{n_i}$, $U_i \in \mathbb{R}^{\bar m_i}$, $W_i \in \mathbb{R}^{\bar p_i}$, $Y_{i} \in \mathbb{R}^{q_{i}}$
	along with a matrix $\mathcal M$ that describes the coupling between the subsystems, with a well-posed interconnection constraint $\mathcal M \prod^{N}_{i=1} Y_{i} \subseteq \mathcal M \prod^{N}_{i=1} W_{i}$. Then the interconnection of subsystems $\Sigma_i, i \in \{1,\ldots,N\}$, denoted by $\mathcal{I}(\Sigma_1,\ldots,\Sigma_N)$, is dt-SS $\Sigma=(X,U,P, \mathcal P, \varsigma,F)$ such that $X := \prod^{N}_{i=1} X_i$, $U := \prod^{N}_{i=1} U_i$, $P:=\prod_{i=1}^{N}P_i$, $\mathcal{P}:=\prod_{i=1}^{N}\mathcal{P}_i$, $\varsigma = [\varsigma_1;\dots;\varsigma_N]$,
	and $F := \prod^{N}_{i=1} F_i$, with the following interconnection constraint:
	\begin{equation}\label{Eq:8}
		[w_1;\ldots ; w_N]= \mathcal M[h_{1};\ldots;h_{N}].
	\end{equation}
	Such an interconnected dt-SS can be represented by
	\begin{equation}\label{Eq_11a}
		\Sigma\!:x(k+1)=f_{\bold{p}(k)}(x(k),\nu(k),\varsigma(k)),
	\end{equation} 
	with $f_p:X\times U \times \mathcal{V}_\varsigma\rightarrow X$.
\end{definition}

\begin{remark}
	Note that the role of $h$ in~\eqref{Eq_1a} is mainly for the sake of interconnecting subsystems as appeared in~\eqref{Eq:8}. Accordingly, the full-state information is assumed to be available for the interconnected system (\emph{i.e.}, its output map is identity) for the sake of controller synthesis.
\end{remark}

In the next section, we present notions of control barrier and storage certificates for, respectively, interconnected dt-SS and individual subsystems.

\section{Control Barrier and Storage Certificates}\label{Sec_CBC}

Here, we first present the notion of control barrier certificates for interconnected dt-SS without disturbance signals. 

\begin{definition}\label{eq:barrier}
	Consider an interconnected system $\Sigma=(X,U,P, \mathcal P, \varsigma,F)$, and $X_{0}, X_{u}\subseteq X$ as, respectively, initial and unsafe sets of the interconnected system. A function $\mathcal B\!:X \times P\rightarrow\mathbb{R}_{\geq0}$ is called a stochastic control barrier certificate (CBC) for $\Sigma$ if, for all $p\in P$,
	\begin{align}\label{eq:B1}
		&\mathcal B(x,p)\leq\gamma, \quad\quad \forall x\in X_0, \\\label{eq:B2}
		&\mathcal B(x,p)\geq\lambda,  \quad\quad \forall x\in X_u,
	\end{align}
	and $\forall x:= x(k)\in X$, $\exists \nu:=\nu(k)\in U$ such that
	\begin{align}\label{eq:B3}
		\sum_{p'=1}^M\pi_{pp'}\EE \Big[\mathcal B(x(k+1),p')\,\big|\,x,\nu,p\Big]  \leq \kappa\mathcal B(x,p) + \psi , 
	\end{align}
	for some $0<\kappa<1$, $\gamma, \lambda,\psi\in\R_{\geq 0}$, with $\lambda > \gamma$, and $M = \Pi_{i=1}^Nm_i$, where $m_i$ is the number of modes for each subsystem $\Sigma_i$ in~\eqref{Eq_1a}.
\end{definition}

The next theorem, adapted from~\cite{1967stochastic}, shows the usefulness of CBC to quantify an upper bound on the probability that the interconnected system reaches certain unsafe regions in finite time horizons.

\begin{theorem}\label{Kushner}
	Let $\Sigma= (X,U,P, \mathcal P, \varsigma,F)$ be an interconnected dt-SS without disturbance signals. Suppose $\mathcal B(x,p)$ is a CBC for $\Sigma$ as in Definition~\ref{eq:barrier}. Then the probability that the solution process of $\Sigma$ starting from any initial state $a\in X_0$ and
	any initial mode $p_0$ reaches $X_u$ under $\nu(\cdot)$ within the time step $k\in [0,\mathcal T]$ is quantified as
	\begin{equation} \label{eqlemma2}
		\PP^{a}_{\nu} \Big\{x^p(k)\in X_u \text{ for some } 0\leq k\leq \mathcal T \, \big|\, x^p(0)=a , p_0\Big\} \leq \delta,
	\end{equation}
	with	
	\begin{equation*}
		\delta\!=  \begin{cases} 
			1-(1-\frac{\gamma}{\lambda})(1-\frac{\psi}{\lambda})^{\mathcal T}, & \quad\quad\quad\text{if } \lambda \geq \frac{\psi}{{1-\kappa}}, \\
			(\frac{\gamma}{\lambda}){\kappa}^{\mathcal T}+(\frac{\psi}{(1-{\kappa})\lambda})(1-{\kappa}^{\mathcal T}), & \quad\quad\quad\text{if } \lambda< \frac{\psi}{{1-\kappa}}.  \\
		\end{cases}
	\end{equation*}
\end{theorem}
\vspace{0.1cm}

In general, finding barrier certificates for large-scale dt-SS as in Definition~\ref{eq:barrier} is computationally very expensive
mainly due to the high dimension of the system. Accordingly, we present in the
following definition a notion of stochastic control \emph{storage} certificates
(CSC) for individual subsystems with disturbance signals
as in~\eqref{Eq_1a}. We then propose in Section~\ref{Sec:Com} our compositional
approach based on dissipativity theory to construct a CBC of the interconnected dt-SS based on CSC of individual subsystems.

\begin{definition}\label{eq:local barrier}
	Consider a dt-SS ~$\Sigma_{p}$, and sets $X_{0}, X_{u}\subseteq X$ as initial and unsafe sets of the subsystem, respectively. A function $\mathcal B_{p}:X\rightarrow\mathbb{R}_{\geq0}$ is called a stochastic control storage certificate (CSC) for $\Sigma_p$ if there exist $0<\kappa_{p}<1$, $\gamma_{p},\lambda_{p}, \psi_{p}\in\R_{\geq 0}$, and a symmetric matrix $\mathcal X$ with conformal block partitions $\mathcal X^{l\bar l}$, $l,\bar l\in\{1,2\}$, such that for all $p\in P$,
	
	\begin{align}\label{eq:LB_c2}
		&\mathcal B_{p}(x)\leq\gamma_{p}, \quad\quad\quad\quad\quad 	\forall x\in X_{0},\\\label{eq:LB_c3}
		&\mathcal B_{p}(x)\geq \lambda_{p}, \quad\quad\quad\quad\quad \forall x\in X_{u},
	\end{align}
	and $\forall x:= x(k)\in X$, $\exists \nu := \nu(k)\in U$ such that $\forall w:=w(k)\in W$,
	\begin{align}\label{eq:B_c4}
		\sum_{p'=1}^m\pi_{pp'}\EE \Big[\mathcal B_{p'}(x(k+1))\,\big|\,x,\nu,w,p\Big]
		\leq \kappa_{p} \mathcal B_{p}(x)+ \psi_{p} + \begin{bmatrix}
			w\\
			h(x)
		\end{bmatrix}^\top\!
		\overbrace{\begin{bmatrix}
				\mathcal X^{11}&\mathcal X^{12}\\
				\mathcal X^{21}&\mathcal X^{22}
		\end{bmatrix}}^{\mathcal X:=}\!\begin{bmatrix}
			w\\
			h(x)
		\end{bmatrix}\!\!.
	\end{align}
\end{definition}
\vspace{0.2cm}

\begin{remark}
	The stochastic storage certificate satisfying conditions~\eqref{eq:LB_c2}-\eqref{eq:B_c4} is useless on its
	own to ensure the safety of the interconnected system. More precisely, stochastic storage certificates are some appropriate tools to construct overall control barrier certificates if some compositionality
	conditions are satisfied (cf.~\eqref{eq:lmi},\eqref{eq:eta}). The safety of the system can then be verified via Theorem~\ref{Kushner}
	only via the constructed \emph{control barrier certificate}.
\end{remark}

In the next section, we analyze networks of stochastic switching subsystems
and show under which conditions one can construct a CBC of an interconnected system using CSC of subsystems.

\section{Compositional Construction of CBC}\label{Sec:Com}

Here, we provide a compositional framework to obtain a CBC of an interconnected dt-SS $\Sigma$ based on CSC of subsystems $\Sigma_i$. Let us assume that there exists a CSC $\mathcal B_{ip_i}$ as in Definition \ref{eq:local barrier} for each subsystem $\Sigma_i, i\in \{1,\dots,N\}$, with  $0<\kappa_{i{p_i}}<1$, $\gamma_{i{p_i}},\lambda_{i{p_i}}, \psi_{i{p_i}}\in\R_{\geq 0}$, and a symmetric matrix $\mathcal X_{i{p_i}}$ with conformal block partitions $\mathcal X_{i{p_i}}^{l\bar l}$, $l,\bar l\in\{1,2\}$. We now propose the following theorem, as the main compositionality result of the work, to provide sufficient conditions for the construction of a CBC of the interconnected dt-SS $\Sigma$ based on CSC of subsystems $\Sigma_i, i \in \{1,\ldots,N\}$. 

\begin{theorem}\label{Thm: Comp}
	Consider an interconnected dt-SS $\Sigma=\mathcal{I}(\Sigma_1,\ldots,$ $\Sigma_N)$ composed of $N$ switching subsystems $\Sigma_i$, $i \in \{1,\ldots,N\}$, with an interconnection matrix $\mathcal M$. Suppose that each switching mode $\Sigma_{i{p}_i}$ admits a CSC $\mathcal B_{i{p}_i}$ as defined in Definition~\ref{eq:local barrier} with initial and unsafe sets $X_{0_i}$ and $X_{u_i}$, respectively. If
	\begin{align}\label{eq:lmi}
		&\begin{bmatrix}
			\mathcal M \\ \mathds{I}_{\tilde q} 
		\end{bmatrix}^\top \!\!\mathcal X_{cmp} \begin{bmatrix}
			\mathcal M \\ \mathds{I}_{\tilde q} 
		\end{bmatrix} \preceq 0,\\\label{eq:eta}
		\sum_{i=1}^N\mu_i& \min_{p_i\in P_i}\{\lambda_{ip_i}\} > \sum_{i=1}^N\mu_i \max_{p_i\in P_i}\{\gamma_{ip_i}\},
	\end{align} 
	where $\mu_i > 0, i \in \{1,\dots,N\}$, then
	\begin{equation} \label{finalCBC}
		\mathcal B(x,p)\Let\sum_{i=1}^N\mu_i\mathcal B_{i{p_i}}(x_i)
	\end{equation}
	with $p =[p_1;\dots;p_N], p_i\in \{1,\dots,m_i\}$, is a CBC for the interconnected system $\Sigma=\mathcal{I}(\Sigma_1,$ $\ldots,\Sigma_N)$, where 
	\begin{equation} \label{xcomp}
		\mathcal X_{cmp} := \begin{bmatrix}
			\mathcal X_1^{11} &  &  &\mathcal X_1^{12} &  &   \\ & \ddots &  &    & \ddots &  & \\
			&    &   \mathcal X_N^{11}  &    &   &  \mathcal X_N^{12} \!\!\!\!\!\!\\
			\mathcal X_1^{21}  &   &   &  \mathcal X_1^{22}  &   &   \\
			& \ddots &  &    & \ddots &  & \\
			&    &   \mathcal X_N^{21}  &    &   &  \mathcal X_N^{22}\!\!\!\!\!\! 
		\end{bmatrix}\!,
	\end{equation}
	and $\tilde q= \sum_{i=1}^N q_{i}$, 
	with $q_{i}$ being the dimension of the output of $\Sigma_i$.  In addition, 
	\begin{align*}
		\gamma & \Let \sum_{i=1}^N\mu_i \max_{p_i\in P_i}\{\gamma_{ip_i}\},~\lambda \Let \sum_{i=1}^N\mu_i \min_{p_i\in P_i}\{\lambda_{ip_i}\},\\
		\kappa s&\Let \max\Big\{\sum_{i=1}^N\mu_i\max_{p_i\in P_i}\{\kappa_{ip_i}\} \mathcal B_{ip_i}(s_i)\big| s_i  {\ge 0},\sum_{i=1}^N \mu_i\mathcal B_{i{p_i}}\!=s\Big\}, \\
		\psi&\Let\sum_{i=1}^N\mu_i\max_{p_i\in P_i}\{\psi_{ip_i}\}.
	\end{align*}
\end{theorem}

\begin{remark}\label{compositionality remark}
	Note that condition \eqref{eq:lmi} is a well-established LMI, discussed in \cite{2016Murat}, as a compositional stability condition based on the dissipativity theory. As shown in \cite{2016Murat}, this condition holds independently of the number of subsystems in many physical applications with particular interconnection topologies (cf. the case study). Condition~\eqref{eq:eta} is not also restrictive since constants $\mu_i$ in~\eqref{finalCBC} play a significant role in rescaling CSC for subsystems while normalizing the effect of gains of other subsystems.
\end{remark}

\section{Computation of CSC and Safety Controller}\label{compute_CBC}

In  this section, we provide an approach based on sum-of-squares (SOS) optimization to compute a CSC and synthesize its corresponding controller for $\Sigma_p$. In order to utilize an SOS optimization, we raise the following assumption.
\begin{assumption}\label{ass:BC}
	Suppose that $\Sigma_p$ has a continuous state set $X\subseteq \mathbb R^{n}$ and continuous input and disturbance sets $U\subseteq \R^{\bar m}$ and $W\subseteq \R^{\bar p}$. Moreover, $f_p:X\times U\times W\times \mathcal{V}_{\varsigma}\rightarrow X$ is a polynomial function of the state $x$ and input and disturbance $\nu, {w}$.
\end{assumption}
Under Assumption~\ref{ass:BC}, the following lemma reformulates conditions~\eqref{eq:LB_c2}-\eqref{eq:B_c4} as an SOS optimization problem.
\begin{lemma}\label{sos}
	Suppose Assumption~\ref{ass:BC} holds and sets $X_0,X_u, X, U, W$ can be defined by vectors of polynomial inequalities as $X_{0}=\{x\in\R^{n}\mid g_{0}(x)\geq0\}$, $X_{u}=\{x\in\R^{n}\mid g_{u}(x)\geq0\}$, $X=\{x\in\R^{n}\mid g(x)\geq0\}$, $U=\{\nu\in\R^{\bar m}\mid g_{\nu}(\nu)\geq0\}$, and $W=\{w\in\R^{\bar p}\mid g_{w}(w)\geq0\}$, where the inequalities are defined element-wise.
	Suppose there exists a sum-of-square polynomial $\mathcal B_{p}(x)$, constants $0<\kappa_{p}<1$, $\gamma_{p},\lambda_{p}, \psi_{p}\in\R_{\geq 0}$, a symmetric matrix $\mathcal X_p$ with conformal block partitions $\mathcal X_p^{l\bar l}$, $l,\bar l\in\{1,2\}$, polynomials $l_{\nu_{j}}(x)$ corresponding to the $j^{\text{th}}$ input in $\nu=(\nu_1,\nu_2,\ldots,\nu_{\bar m})\in U\subseteq \R^{\bar m}$, and vectors of sum-of-squares polynomials $l_{0_{p}}(x)$, $l_{u_{p}}(x)$, $l_{p}(x,\nu,w)$, $l_{\nu_{p}}(x,\nu,w)$, and $l_{w_{p}}(x,\nu,w)$ of appropriate dimensions such that the following expressions are sum-of-squares polynomials, for all $p\in P$:
	\begin{align}\label{eq:sos1}
		-&\mathcal B_{p}(x)-l_{0_{p}}^\top(x) g_{0}(x)+\gamma_{p}\\\label{eq:sos2}
		&\mathcal B_{p}(x)-l_{u_{p}}^\top(x) g_{u}(x)-\lambda_{p} \\\notag
		-&\sum_{p'=1}^m\pi_{pp'}\EE \Big[\mathcal B_p'(x(k+1))\,\big|\,x,\nu,w,p\Big] + \kappa_{p} \mathcal B_{p}(x)+ \psi_{p}+ \begin{bmatrix}
		w\\
		h(x)
		\end{bmatrix}^\top\!
		\begin{bmatrix}
		\mathcal X^{11}&\mathcal X^{12}\\
		\mathcal X^{21}&\mathcal X^{22}
		\end{bmatrix}\!\begin{bmatrix}
		w\\
		h(x)
		\end{bmatrix}\\\label{eq:sos3}
		& - \sum_{j=1}^{\bar m}(\nu_j-l_{\nu_{j_p}}(x))-l_{p}^\top(x,\nu,w) g(x)-l_{w_{p}}^\top(x,\nu,w) g_{w}(w)-l_{\nu_{p}}^\top(x,\nu,w) g_{\nu}(\nu).
	\end{align}
	Then, $\mathcal B_p(x)$ satisfies conditions~\eqref{eq:LB_c2}-\eqref{eq:B_c4} in Definition~\ref{eq:local barrier} and $\nu=[l_{\nu_{{1}_p}}(x);\dots;l_{\nu_{{\bar m}_p}}(x)]$ is the corresponding controller employed at the mode $p\in P$.
\end{lemma}

\section{Case Study: Room Temperature Network}\label{Sec:Case}

To demonstrate the effectiveness of our proposed results, we apply them to a room temperature network in a circular building containing $N = 200$ rooms. The model of this case study is borrowed from~\cite{Meyer.2018} by including the stochasticity in the model as an additive noise. The evolution of the temperature in the interconnected system is governed by the following stochastic switching system:
\begin{equation*}
	\Sigma:
	x(k+1)=Ax(k) + \beta T_h\nu(k) + \alpha T_{\bold{p}(k)} + R_{\bold{p}(k)}\varsigma(k),
\end{equation*}
where $A \in \mathbb{R}^{n \times n}$ is a matrix with diagonal elements of $\bar a_{ii}=(1-2 \theta-\alpha-\beta\nu_i(k))$, off-diagonal elements $\bar a_{i,i+1}=\bar a_{i+1,i}=\bar a_{1,n}=\bar a_{n,1}= \theta$, $i\in \{1,\ldots,N-1\}$, and all other elements being identically zero. Parameters $\theta = 0.005$, $\alpha = 0.06$, and $\beta=0.145$ are conduction factors, respectively, between rooms $i \pm 1$ and $i$, the external environment and the room $i$, and the heater and the room $i$.  Outside temperatures are the same for all rooms: $T_{\bold{p}(k)}=\bar T_{p_i}\mathds{I}_n$ with $\bar T_{p_i} =\left\{\hspace{-1.7mm}\begin{array}{l} -15\,{}^{\circ}\mathsf{C},\quad\text{if}~~~  p_i = 1,\\
-20\,{}^{\circ}\mathsf{C},\quad\text{if}~~~  p_i = 2.\\
\end{array}\right.$ The heater temperature is $T_h=45\,{}^{\circ}\mathsf{C}$, and $R_{\bold{p}(k)} = \bar R_{p_i}\mathds{I}_n$ with $\bar R_{p_i} =\left\{\hspace{-1.7mm}\begin{array}{l} 0.3,\quad\text{if}~~~  p_i = 1,\\
0.5,\quad\text{if}~~~  p_i = 2.\\
\end{array}\right.$ Moreover, $ x(k)=[x_1(k);\ldots;x_N(k)]$, $\nu(k)=[\nu_1(k);\ldots;\nu_N(k)]$, and  $\varsigma=[\varsigma_1(k);\ldots;\varsigma_N(k)]$.

Now by considering individual rooms as $\Sigma_i$ represented by
\begin{align}\notag
	\Sigma_i\!: \begin{cases}
		x_i(k+1) = \bar a_{ii}x_i(k) + \beta_i T_h \nu_i(k) +  \theta_i w_i(k) + \alpha_i \bar T_{ip_i} + \bar R_{ip_i} \varsigma_i(k), \\
		y_i(k)=x_i(k),
	\end{cases}
\end{align}
one can readily verify that $\Sigma=\mathcal{I}(\Sigma_1,\ldots,\Sigma_N)$, where the coupling matrix $\mathcal M$ is defined as $\hat m_{i,i+1}=\hat m_{i+1,i}=\hat m_{1,N}=\hat m_{N,1}=1$, $i\in \{1,\ldots,N-1\}$, and all other elements are identically zero. 

Transition probability matrix for switching between two modes is given as $\pi=\begin{bmatrix}0.3 & 0.7 \\ 0.4  & 0.6 \end{bmatrix}$\!.  
In addition, the regions of interest are given as $X_i \in [1,50], X_{0_i} \in [19.5,20], X_{u_i} = [1,17]\cup [23,50], \forall i\in\{1,\dots,200\}$. The main goal is to find a CBC for the interconnected system and its corresponding safety controller such that the temperature of rooms remains in the comfort zone $[17,23]^{200}$. To do so, we first search for CSC and accordingly design local controllers for subsystems $\Sigma_i$. Consequently, the controller for the interconnected system $\Sigma$ would be a vector such that each of its components is a controller for subsystems $\Sigma_i$. 

We employ the software tool \textsf{SOSTOOLS}~\cite{papachristodoulou2013sostools} and the SDP solver \textsf{SeDuMi}~\cite{sturm1999using} to compute CSC as described in Section~\ref{compute_CBC}. Based on Lemma~\ref{sos}, we compute CSC of an order $4$ as $\mathcal B_{ip_i}(x_i) = 0.00242x_i^4 - 0.091x_i^3 + 0.7696x_i^2 + 1.4935 x_i + 3.1329$ and the corresponding controller $\nu_{ip_i} = - 0.0121x_i + 0.8$ for $p_i = 1$, and $\mathcal B_{ip_i}(x_i) = 0.00191x_i^4 - 0.0718x_i^3 + 0.5998x_i^2 + 1.2424x_i + 3.2433$ together with $\nu_{ip_i} = - 0.02527x_i + 1.15$ for $p_i = 2, \forall i\in\{1,\dots,200\}$. Moreover, the corresponding constants in Definition~\ref{eq:local barrier} satisfying conditions~\eqref{eq:LB_c2}-\eqref{eq:B_c4} are quantified as $\gamma_{ip_i} = 0.13, \lambda_{ip_i} = 4.4, \kappa_{ip_i} = 0.91, \psi_{ip_i} = 0.001$ for $p_i = 1$, and $\gamma_{ip_i} = 0.14, \lambda_{ip_i} = 4.3, \kappa_{ip_i} = 0.92, \psi_{ip_i} = 0.0015$ for $p_i = 2, \forall i\in\{1,\dots,200\}$, and
\begin{align}\label{Eq_221}
	\mathcal X_i=\begin{bmatrix} 0.005 & 0.003  \\ 0.003  &  -0.035 \end{bmatrix}\!\!.
\end{align}
We now proceed with Theorem~\ref{Thm: Comp} to construct a CBC for the interconnected system using CSC of subsystems. By selecting $\mu_i = 1, \forall i\in\{1,\dots,200\}$, and utilizing $\mathcal X_i$ in (\ref{Eq_221}), the matrix $\mathcal X_{cmp}$ in \eqref{xcomp} is reduced to
$$
\mathcal X_{cmp}=\begin{bmatrix} 0.005~\!\mathds{I}_{200} & 0.003~\!\mathds{I}_{200} \\ 0.003~\!\mathds{I}_{200} & -0.035~\!\mathds{I}_{200} \end{bmatrix}\!\!,
$$
and condition \eqref{eq:lmi} is reduced to
\begin{align*}
	\begin{bmatrix} \mathcal M  \\ \mathds{I}_{200} \end{bmatrix}^\top\!\!\mathcal X_{cmp}\begin{bmatrix} \mathcal M \\ \mathds{I}_{200} \end{bmatrix} =~\!& 0.005\mathcal M^\top\mathcal M- 0.003 (\mathcal M + \mathcal M^\top\!)-0.035~\!\mathds{I}_{200} \preceq 0,
\end{align*}
which is always satisfied without requiring any restrictions on the number or gains of subsystems. To show this, we employed the property of the interconnection topology as $\mathcal M=\mathcal M^\top$\! and Gershgorin circle theorem \cite{bell1965gershgorin}. Moreover, the compositionality condition~\eqref{eq:eta} is also met since $\sum_{i=1}^{200}\mu_i \min_{p_i\in P_i}\{\lambda_{ip_i}\} > \sum_{i=1}^{200}\mu_i \max_{p_i\in P_i}\{\gamma_{ip_i}\}$. Then by employing the results of Theorem~\ref{Thm: Comp}, one can conclude that $\mathcal B(x,p)\Let\sum_{i=1}^{200}\mathcal B_{i{p_i}}(x_i)$ is a CBC for the interconnected system with $\gamma = \sum_{i=1}^{200}\mu_i \max_{p_i\in P_i}\{\gamma_{ip_i}\} = 28, \lambda = \sum_{i=1}^{200}\mu_i\\ \min_{p_i\in P_i}\{\lambda_{ip_i}\} = 860, \kappa = 0.92,$ and $\psi = \sum_{i=1}^{200}\max_{p_i\in P_i}\{\psi_{ip_i}\} = 0.3$. 

By employing Theorem~\ref{Kushner}, we guarantee that the temperature of the interconnected system $\Sigma$ starting from initial conditions inside $X_{0} = [19.5~20]^{200}$ remains in the safe set $[17~23]^{200}$ during the time horizon $\mathcal T=100$ with the probability of at least $94\%$, \emph{i.e.,}
\begin{align}\label{threshold}
	&\PP^{a}_{\nu} \Big\{x^p(k)\notin X_u \, \big|\, x^p(0)=a , p_0, \forall k\in[0,100]\Big\} \geq 0.94.
\end{align}

\begin{figure}
	\centering
	\includegraphics[width=7.7cm]{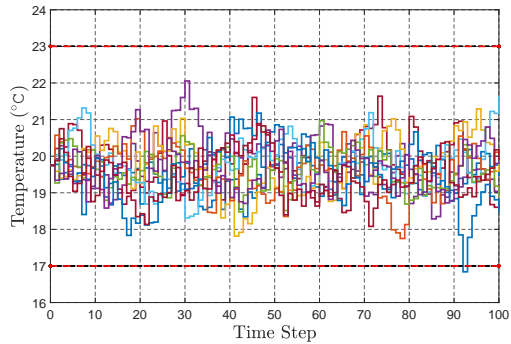}
	\caption{Closed-loop state trajectories of a representative room with $10$ noise realizations in a network of $200$ rooms for $\mathcal T = 100$.}
	\label{Simulation}
\end{figure}

Closed-loop state trajectories of a representative room with $10$ different noise realizations are illustrated in Fig.~\ref{Simulation}. The computation of CSC and its corresponding controller for each individual room took almost $25$ seconds with a memory usage of $1.4$ MB on a Windows operating system (Intel i7@3.6GHz CPU and 32 GB of RAM). It is worth mentioning that in order
to empirically verify the proposed probabilistic
bound in~\eqref{threshold}, one can run Monte Carlo simulations with a high number of noise realizations.

\section{Conclusion}
In this work, we proposed a compositional framework based on dissipativity theory for the safety controller synthesis of large-scale discrete-time stochastic systems with \emph{Markovian switching signals}. We first introduced the notion of control storage certificates, using which one can construct control barrier certificates of interconnected
systems by leveraging dissipativity-type compositionality
conditions. We then provided upper bounds on the probability
that interconnected systems reach unsafe regions
in finite time horizons. We formulated our proposed conditions
to a sum-of-squares optimization problem to systematically
search for storage certificates and
corresponding local controllers enforcing safety properties. We finally verified our results on a room temperature network of $200$ individual rooms while admitting Markovian switching signals.

\bibliographystyle{IEEEtran}
\bibliography{biblio}

\begin{thebibliography}{10}
\providecommand{\url}[1]{#1}
\csname url@samestyle\endcsname
\providecommand{\newblock}{\relax}
\providecommand{\bibinfo}[2]{#2}
\providecommand{\BIBentrySTDinterwordspacing}{\spaceskip=0pt\relax}
\providecommand{\BIBentryALTinterwordstretchfactor}{4}
\providecommand{\BIBentryALTinterwordspacing}{\spaceskip=\fontdimen2\font plus
\BIBentryALTinterwordstretchfactor\fontdimen3\font minus
  \fontdimen4\font\relax}
\providecommand{\BIBforeignlanguage}[2]{{%
\expandafter\ifx\csname l@#1\endcsname\relax
\typeout{** WARNING: IEEEtran.bst: No hyphenation pattern has been}%
\typeout{** loaded for the language `#1'. Using the pattern for}%
\typeout{** the default language instead.}%
\else
\language=\csname l@#1\endcsname
\fi
#2}}
\providecommand{\BIBdecl}{\relax}
\BIBdecl

\bibitem{pnueli1977temporal}
A.~Pnueli, ``The temporal logic of programs,'' in \emph{Proceedings of the 18th
  Annual Symposium on Foundations of Computer Science}, 1977, pp. 46--57.

\bibitem{julius2009approximations}
A.~A. Julius and G.~J. Pappas, ``Approximations of stochastic hybrid systems,''
  \emph{IEEE Transactions on Automatic Control}, vol.~54, no.~6, pp.
  1193--1203, 2009.

\bibitem{APLS08}
A.~Abate, M.~Prandini, J.~Lygeros, and S.~Sastry, ``Probabilistic reachability
  and safety for controlled discrete-time stochastic hybrid systems,''
  \emph{Automatica}, vol.~44, no.~11, pp. 2724--2734, 2008.

\bibitem{zamani2015symbolic}
M.~Zamani, A.~Abate, and A.~Girard, ``Symbolic models for stochastic switched
  systems: A discretization and a discretization-free approach,''
  \emph{Automatica}, vol.~55, pp. 183--196, 2015.

\bibitem{zamani2014symbolic}
M.~Zamani, P.~Mohajerin~Esfahani, R.~Majumdar, A.~Abate, and J.~Lygeros,
  ``Symbolic control of stochastic systems via approximately bisimilar finite
  abstractions,'' \emph{IEEE Transactions on Automatic Control}, vol.~59,
  no.~12, pp. 3135--3150, 2014.

\bibitem{tmka2013}
I.~Tkachev, A.~Mereacre, J.-P. Katoen, and A.~Abate, ``Quantitative
  automata-based controller synthesis for non-autonomous stochastic hybrid
  systems,'' in \emph{Proceedings of the 16th ACM International Conference on
  Hybrid Systems: Computation and Control}, 2013, pp. 293--302.

\bibitem{SAM17}
S.~Soudjani, A.~Abate, and R.~Majumdar, ``Dynamic {B}ayesian networks for
  formal verification of structured stochastic processes,'' \emph{Acta
  Informatica}, vol.~54, no.~2, pp. 217--242, 2017.

\bibitem{hahn2013compositional}
E.~M. Hahn, A.~Hartmanns, H.~Hermanns, and J.-P. Katoen, ``A compositional
  modelling and analysis framework for stochastic hybrid systems,''
  \emph{Formal Methods in System Design}, vol.~43, no.~2, pp. 191--232, 2013.

\bibitem{lavaei2018CDCJ}
A.~Lavaei, S.~Soudjani, and M.~Zamani, ``Compositional construction of infinite
  abstractions for networks of stochastic control systems,'' \emph{Automatica},
  vol. 107, pp. 125--137, 2019.

\bibitem{lavaei2019HSCC_J}
------, ``Compositional abstraction-based synthesis for networks of stochastic
  switched systems,'' \emph{Automatica}, vol. 114, 2020.

\bibitem{lavaei2018ADHSJ}
------, ``Compositional (in)finite abstractions for large-scale interconnected
  stochastic systems,'' \emph{IEEE Transactions on Automatic Control}, vol.~65,
  no.~12, pp. 5280--5295, 2020.

\bibitem{Lavaei_TAC2022}
A.~Lavaei and M.~Zamani, ``From dissipativity theory to compositional synthesis
  of large-scale stochastic switched systems,'' \emph{IEEE Transactions on
  Automatic Control}, 2022.

\bibitem{Lavaei_Survey}
A.~Lavaei, S.~Soudjani, A.~Abate, and M.~Zamani, ``Automated verification and
  synthesis of stochastic hybrid systems: A survey,'' \emph{Automatica}, 2022.

\bibitem{AmyJournal2020}
A.~Nejati, S.~{Soudjani}, and M.~Zamani, ``Compositional abstraction-based
  synthesis for continuous-time stochastic hybrid systems,'' \emph{European
  Journal of Control}, vol.~57, pp. 82--94, 2021.

\bibitem{prajna2004safety}
S.~Prajna and A.~Jadbabaie, ``Safety verification of hybrid systems using
  barrier certificates,'' in \emph{Proceedings of the International Workshop on
  Hybrid Systems: Computation and Control (HSCC)}, 2004, pp. 477--492.

\bibitem{wieland2007constructive}
P.~Wieland and F.~Allg{\"o}wer, ``Constructive safety using control barrier
  functions,'' \emph{IFAC Proceedings Volumes}, vol.~40, no.~12, pp. 462--467,
  2007.

\bibitem{prajna2007framework}
S.~Prajna, A.~Jadbabaie, and G.~J. Pappas, ``A framework for worst-case and
  stochastic safety verification using barrier certificates,'' \emph{IEEE
  Transactions on Automatic Control}, vol.~52, no.~8, pp. 1415--1428, 2007.

\bibitem{wisniewski2017stochastic}
R.~Wisniewski and M.~L. Bujorianu, ``Stochastic safety analysis of stochastic
  hybrid systems,'' in \emph{Proceedings of the 56th IEEE Conference on
  Decision and Control}, 2017, pp. 2390--2395.

\bibitem{huang2017probabilistic}
C.~Huang, X.~Chen, W.~Lin, Z.~Yang, and X.~Li, ``Probabilistic safety
  verification of stochastic hybrid systems using barrier certificates,''
  \emph{ACM Transactions on Embedded Computing Systems (TECS)}, vol.~16,
  no.~5s, p. 186, 2017.

\bibitem{Pushpak2019}
P.~{Jagtap}, S.~{Soudjani}, and M.~{Zamani}, ``Formal synthesis of stochastic
  systems via control barrier certificates,'' \emph{IEEE Transactions on
  Automatic Control}, 2020.

\bibitem{clark2019control}
A.~Clark, ``Control barrier functions for complete and incomplete information
  stochastic systems,'' in \emph{Proceedings of the American Control Conference
  (ACC)}, 2019, pp. 2928--2935.

\bibitem{santoyo2019verification}
C.~Santoyo, M.~Dutreix, and S.~Coogan, ``Verification and control for
  finite-time safety of stochastic systems via barrier functions,'' in
  \emph{Proceedings of the IEEE Conference on Control Technology and
  Applications}, 2019, pp. 712--717.

\bibitem{ahmadi2019safe}
M.~Ahmadi, A.~Singletary, J.~W. Burdick, and A.~D. Ames, ``Safe policy
  synthesis in multi-agent {POMDPs} via discrete-time barrier functions,'' in
  \emph{Proceedings of the 58th Conference on Decision and Control (CDC)},
  2019, pp. 4797--4803.

\bibitem{Niloofar_TNCS_2022}
N.~Jahanshahi, A.~Lavaei, and M.~Zamani, ``Compositional construction of safety
  controllers for networks of continuous-space pomdps,'' \emph{IEEE
  Transactions on Control of Network Systems}, 2022.

\bibitem{ames2019control}
A.~D. Ames, S.~Coogan, M.~Egerstedt, G.~Notomista, K.~Sreenath, and P.~Tabuada,
  ``Control barrier functions: Theory and applications,'' in \emph{Proceedings
  of the 18th European Control Conference (ECC)}, 2019, pp. 3420--3431.

\bibitem{LavaeiIFAC2020}
M.~Anand, A.~Lavaei, and M.~Zamani, ``Compositional construction of control
  barrier certificates for large-scale interconnected stochastic systems,''
  \emph{Proceedings of the 21st IFAC World Congress}, vol.~53, no.~2, pp.
  1862--1867, 2020.

\bibitem{AmyIFAC12020}
A.~Nejati, S.~{Soudjani}, and M.~Zamani, ``Compositional construction of
  control barrier functions for networks of continuous-time stochastic
  systems,'' \emph{Proceedings of the 21st IFAC World Congress}, vol.~53,
  no.~2, pp. 1856--1861, 2020.

\bibitem{Amy_LCSS20}
------, ``Compositional construction of control barrier certificates for
  large-scale stochastic switched systems,'' \emph{IEEE Control Systems
  Letters}, vol.~4, no.~4, pp. 845--850, 2020.

\bibitem{puterman2014markov}
M.~L. Puterman, \emph{Markov decision processes: discrete stochastic dynamic
  programming}.\hskip 1em plus 0.5em minus 0.4em\relax John Wiley \& Sons,
  2014.

\bibitem{du2021certainty}
Z.~Du, Y.~Sattar, D.~A. Tarzanagh, L.~Balzano, S.~Oymak, and N.~Ozay,
  ``Certainty equivalent quadratic control for markov jump systems,''
  \emph{arXiv:2105.12358}, 2021.

\bibitem{1967stochastic}
H.~J. Kushner, \emph{Stochastic Stability and Control}, ser. Mathematics in
  Science and Engineering.\hskip 1em plus 0.5em minus 0.4em\relax Elsevier
  Science, 1967.

\bibitem{2016Murat}
M.~Arcak, C.~Meissen, and A.~Packard, \emph{Networks of dissipative systems},
  ser. SpringerBriefs in Electrical and Computer Engineering.\hskip 1em plus
  0.5em minus 0.4em\relax Springer, 2016.

\bibitem{Meyer.2018}
P.-J. Meyer, A.~Girard, and E.~Witrant, ``Compositional abstraction and safety
  synthesis using overlapping symbolic models,'' \emph{IEEE Transactions on
  Automatic Control}, vol.~63, no.~6, pp. 1835--1841, 2018.

\bibitem{papachristodoulou2013sostools}
A.~Papachristodoulou, J.~Anderson, G.~Valmorbida, S.~Prajna, P.~Seiler, and
  P.~Parrilo, ``{SOSTOOLS} version 3.00 sum of squares optimization toolbox for
  {MATLAB},'' \emph{arXiv:1310.4716}, 2013.

\bibitem{sturm1999using}
J.~F. Sturm, ``Using {SeDuMi} 1.02, a {MATLAB} toolbox for optimization over
  symmetric cones,'' \emph{Optimization methods and software}, vol.~11, no.
  1-4, pp. 625--653, 1999.

\bibitem{bell1965gershgorin}
H.~E. Bell, ``Gershgorin's theorem and the zeros of polynomials,'' \emph{The
  American Mathematical Monthly}, vol.~72, no.~3, pp. 292--295, 1965.

\end{thebibliography}

\end{document}